\documentclass[pra,aps,showpacs,groupedaddress,superscriptaddress,twocolumn, numerical]{revtex4-1}

\usepackage[utf8x]{inputenc}
\usepackage{color}
\usepackage{bbm} 

\usepackage{nicefrac}
\usepackage{amsfonts,amsmath,amssymb,stmaryrd}
\usepackage{braket}

 \usepackage[autostyle=true]{csquotes}

\usepackage{tabularx}
\usepackage{multirow} 
\usepackage{hhline}
\usepackage{graphicx}
\usepackage{subfigure}  
\usepackage{bbm} 
\usepackage{hyperref}
\usepackage[pdftex]{epsfig}
\usepackage{mathrsfs}
\usepackage{verbatim}
\usepackage{centernot}
\usepackage{ulem}
\usepackage{array}
\usepackage{cancel}
\usepackage{ifthen}
\usepackage{bm}	
\usepackage{siunitx}
\usepackage{todonotes}
\InputIfFileExists{gitinfo-latex.inc}{}{}

\newcounter{lastnote}

\begin{document}
\title{Nonergodic Diffusion of Single Atoms in a Periodic Potential}
\author{Farina Kindermann}
\affiliation{Department of Physics and Research Center OPTIMAS, University of Kaiserslautern, Germany}

\author{Andreas Dechant}
\affiliation{Department of Physics, Friedrich Alexander University Erlangen-N\"urnberg, Germany}
\affiliation{Department of Physics, Kyoto University, Japan}

\author{Michael Hohmann}
\affiliation{Department of Physics and Research Center OPTIMAS, University of Kaiserslautern, Germany}

\author{Tobias Lausch}
\affiliation{Department of Physics and Research Center OPTIMAS, University of Kaiserslautern, Germany}

\author{Daniel Mayer}
\affiliation{Department of Physics and Research Center OPTIMAS, University of Kaiserslautern, Germany}
\affiliation{Graduate School Materials Science in Mainz, Gottlieb-Daimler-Strasse 47, 67663 Kaiserslautern, Germany}

\author{Felix Schmidt}
\affiliation{Department of Physics and Research Center OPTIMAS, University of Kaiserslautern, Germany}
\affiliation{Graduate School Materials Science in Mainz, Gottlieb-Daimler-Strasse 47, 67663 Kaiserslautern, Germany}

\author{Eric Lutz}
\affiliation{Department of Physics, Friedrich Alexander University Erlangen-N\"urnberg, Germany}

\author{Artur Widera}
\affiliation{Department of Physics and Research Center OPTIMAS, University of Kaiserslautern, Germany}
\affiliation{Graduate School Materials Science in Mainz, Gottlieb-Daimler-Strasse 47, 67663 Kaiserslautern, Germany}

\date{\today}

\begin{abstract}
Diffusion is a central phenomenon in almost all fields of natural science revealing microscopic processes from the observation of macroscopic dynamics. 
Here, we consider the paradigmatic system of a single atom diffusing in a periodic potential. We engineer microscopic particle-environment interaction to control the ensuing diffusion over a broad range of diffusion constants and from normal to subdiffusion. 
While one- and two-point properties extracted from single particle trajectories, such as variance or position correlations, indicate apparent Brownian motion, the step size distribution, however, shows exponentially decaying tails. 
Furthermore non-ergodic dynamics is observed on long time scales. 
We demonstrate excellent agreement with a model of continuous time random walk with exponential distribution, which applies to various transport phenomena in condensed or soft matter with periodic potentials.
\end{abstract}	

\maketitle

\begin{figure}
\includegraphics[width=0.49\textwidth]{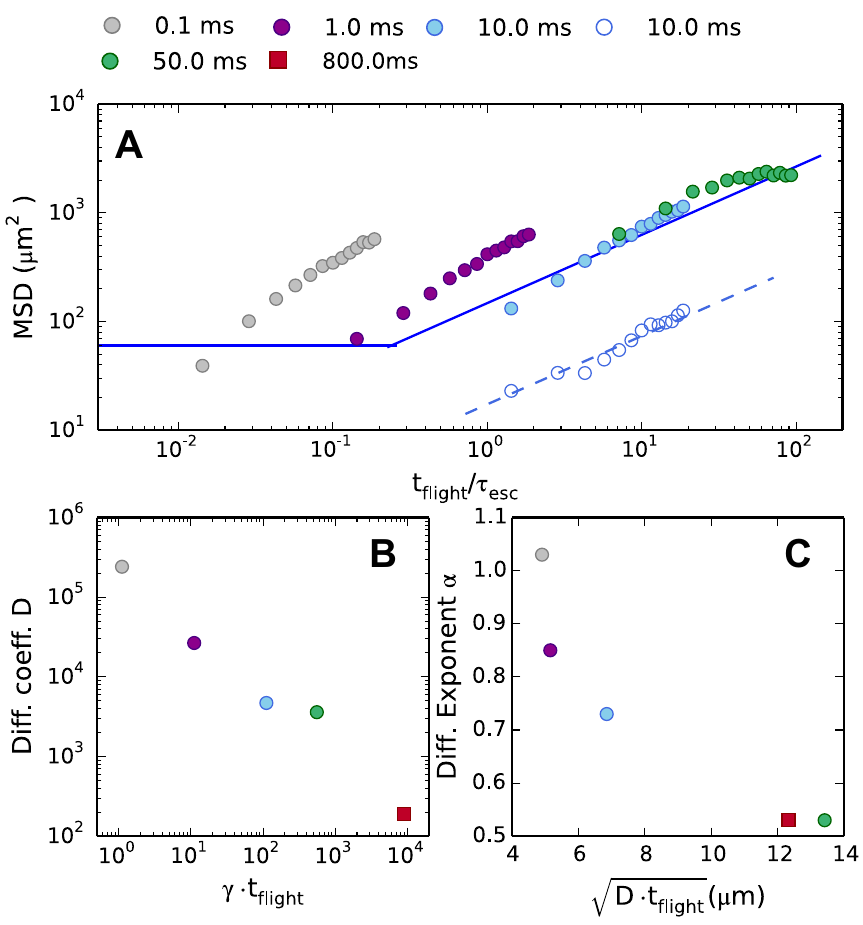}
\caption{(A) MSDs for different $t_\mathrm{flight}$ between each image (denoted by the symbols in the legend). Filled (open) symbols are with (without) molasses present for a potential depth $U_\mathrm{low}=\SI{210}{\micro \K}$.
For $t_\mathrm{flight} \cdot \gamma \ll 1$, the diffusion is dominated by heating phases leading to enhanced diffusion. For longer $t_\mathrm{flight}$ the cooling is significant and the diffusion is essentially slowed down. (B) The extracted diffusion coefficient decreases for longer flight times as the atoms experience larger characteristic flight distances and hence the confining dipole trap potential leads to more and more subdiffussive behaviour as indicated by the diffusion exponent shown in (C). 
}
\label{fig:diffusion}
\end{figure}

The concept of diffusion  is ubiquitous. It does not only plays a prominent role in physics \cite{Frey2005}, chemistry \cite{Kramers1940} and biology \cite{Jeon2011}, but also in economics and finance \cite{Lax2001}, as well as in vehicular traffic \cite{Kerner2004}.  Recent developments have lead to a better understanding of the diffusive properties of increasingly complex particles,  from underdamped colloid particles  \cite{Li2010} and  anisotropic ellipsoids \cite{Han2006} to extended stiff filaments \cite{Fakhri2010} and  fluidized matter \cite{DAnna2003}. On the  other hand, the diffusion of tracer particles has become  a powerful tool to probe the properties of complex systems from turbulent fluids \cite{Porta2001}  to living cells \cite{Jeon2011}.
The hallmarks of standard Brownian diffusion \cite{Lax2001} are (i)  a linear mean-square displacement (MSD), $\sigma^2(t) = \langle\Delta x_t^2 \rangle- \langle \Delta x_t\rangle^2  \propto 2Dt$, where   $D$ is the diffusion coefficient and $\langle \cdot \rangle$  denotes the average  over many trajectories, (ii) a Gaussian step-size distribution,   a direct consequence of the central-limit theorem, and (iii) ergodic behavior in a potential, implying that ensemble and time averages are equal in the long-time limit. The ergodic property lies at the core of statistical physics and indicates that a single trajectory is representative for the ensemble. However, an increasing number of systems exhibit nonergodic features owing to slow, nonexponential relaxation. Examples include blinking quantum dots \cite{Brokmann2002},  the motion of lipid granules \cite{Jeon2011},  of mRNA molecules \cite{He2008} and of receptors in living cells \cite{Manzo2014}. These systems lie outside the range of standard statistical physics. Important questions for their understanding are  the time scale at which thermalization and ergodicity are established, as well as the microscopic mechanisms that prevent ergodicity on long time scales.
Often in experiments only the ensemble averaged MSD is measured from which trajectory properties, such as ergodicity, cannot be determined. It has however been shown that the MSD does not unambiguously characterize the microscopic dynamics \cite{Leptos2009,Meroz2011}. For example, non-apparent Brownian motion with a linear MSD but non-Gaussian probability distributions were observed for colloids diffusing on soft polymers \cite{Wang2009,Wang2012}.
 Here we experimentally realize an ideal system of a single atom diffusing in a periodic potential \cite{Risken1978}. This paradigmatic model has been extensively used to describe for instance superionic conductors, phase-locked loops and surfaces \cite{Fulde1975,viterbi2013,Sancho2004}.
We demonstrate experimentally as well as theoretically that this system intrinsically exhibits Brownian properties for one-time and two-time properties, for example variance and position correlation, extracted from single-particle trajectories. The central limit theorem, in contrast, does not apply even for times much larger than all characteristic time scales in the system. This leads to deviation of the step size distribution from the Brownian case and also to nonergodic dynamics. Importantly, neither the MSD nor the position correlation function in our measurements feature any signal of such markedly different behaviour.
Furthermore, exploiting our exceptional external control over experimental parameters we engineer the diffusion to become increasingly subdiffusive in order to follow the non-Brownian character of the dynamics.
Experimentally, we use a single ultracold Caesium (Cs) atom trapped in a periodic potential and interacting with a near resonant light field acting as a bath. 
This system realizes the most elementary and thus controllable diffusion process of a single particle without uncontrolled internal degrees of freedom. We use single rather than ensembles of atoms, to unambiguously identify individual particle trajectories.
By adjusting external properties of the laser beams interacting with the atom, such as intensity or frequency, parameters of the diffusion, e.g. the diffusion constant, damping or temperature of the bath, can be tightly controlled. 
\begin{figure}
\includegraphics[width=0.4\textwidth]{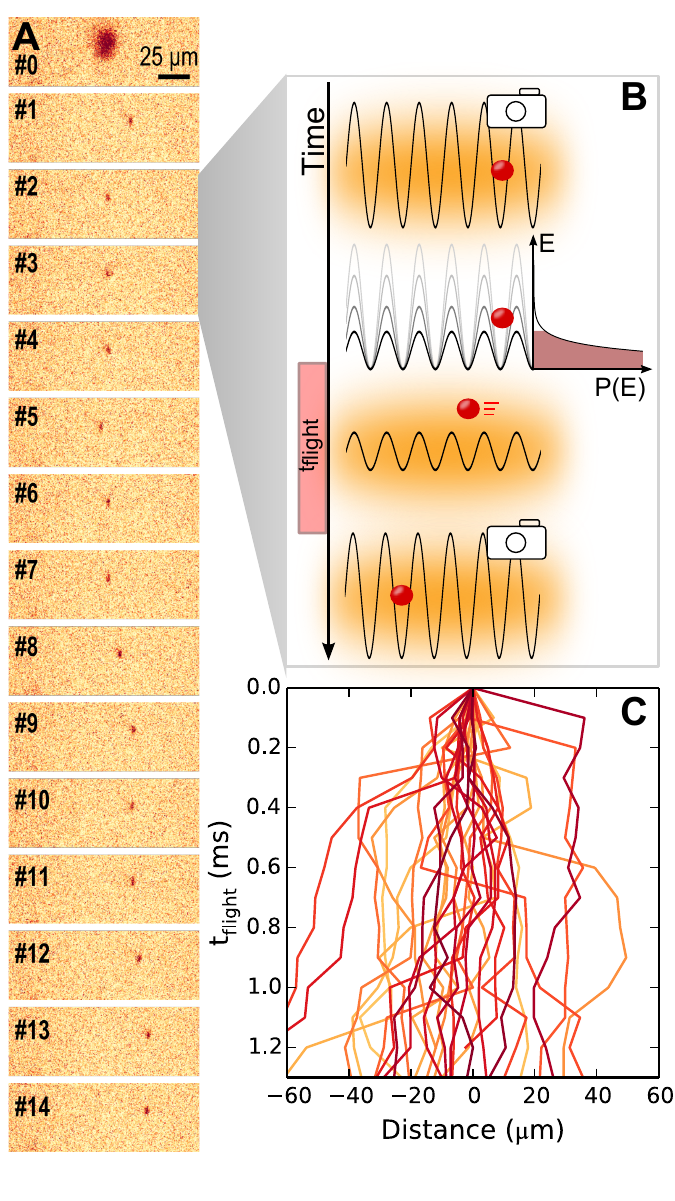}
\caption{(A) Series of fluorescence images  from a single atom. In the magneto optical trap the atom number is deduced (image \#0).
The next 14 images are taken in the lattice and reveal the atoms position. 
(B) Schematic of the sequence. After imaging acquisition, the molasses intensity is reduced to zero and the lattice potential is adiabatically lowered to a value $U_\mathrm{low}$ to ensue thermal classical hopping.
Only atoms with energies larger than the potential barrier contribute the the diffusion.
Switching on the molasses initiates the atom to move and the diffusion evolves for a flight time $t_\mathrm{flight}$ between \SI{0.1}{\milli \second} and \SI{50}{\milli \second}.
The lattice potential is ramped up instantaneously for the next image process to freeze the atoms position.
The process is repeated after each image acquisition.
(C) A typical set of 30 single atom traces for $t_\mathrm{flight} = \SI{0.1}{\milli \second}$.
}
\label{fig:setup}
\end{figure}
In order to initialize the system, we prepare a single laser-cooled atom, thermalized with a laser light field at a temperature of $T\approx \SI{50}{\micro \K}$. We trap the particle in a periodic potential of a 1D optical lattice with depth $U_0 = k_B \times \SI{850}{\micro \K}$ and lattice spacing of $\lambda/2 = \SI{395}{\nano \m}$, where no dynamics is observed. Here $k_B$  is the Boltzmann constant and $\lambda$ the wavelength of the lattice laser light.
Along the lattice axis, the atom is trapped in the nodes of the interference light field.
Transversely the atom is confined by a \SI{1064}{\nano \m} running wave dipole trap, spatially overlapped with the lattice axis. The dipole trap contributes a harmonic confinement along the lattice axis with trapping frequency $\omega \approx 2\pi\times 60\,$Hz for atoms leaving the lattice potential, which will be important to control the character of the diffusion. Hence, the atom is confined to disc-shaped lattice wells, performing rapid in-well dynamics, but diffusion dynamics along the the lattice axis is the main process. This diffusion dynamics is initiated by, first, adiabatically lowering the depth to $U_{low} = k_B \times \SI{210}{\micro \K}$; and, second, illuminating the atom with near resonant light. 
After a variable interaction time $t_\mathrm{flight}$ we freeze the atomic motion by rapidly increasing the lattice depth to $U_0$ again.
The atomic position is detected from high resolution fluorescence images, see Figure \ref{fig:setup}A and \ref{fig:setup}B, before the potential depth is lowered for a next diffusion step.
We stroboscopically image the same atom 14 times, and we record $600 \ldots 1000$ of such traces for each parameter set. Importantly, taking an image involves scattering of approximately $10^6$ photons off the tightly confined atom, so that all properties of the previous diffusion step are effectively reset and no memory on, e.g., velocity or temperature can be retained. This is a requirement of a Markov process, meaning that any step of the random walk only depends on the previous one and not on the complete history of the walk.
\begin{figure}
\includegraphics[width=0.49\textwidth]{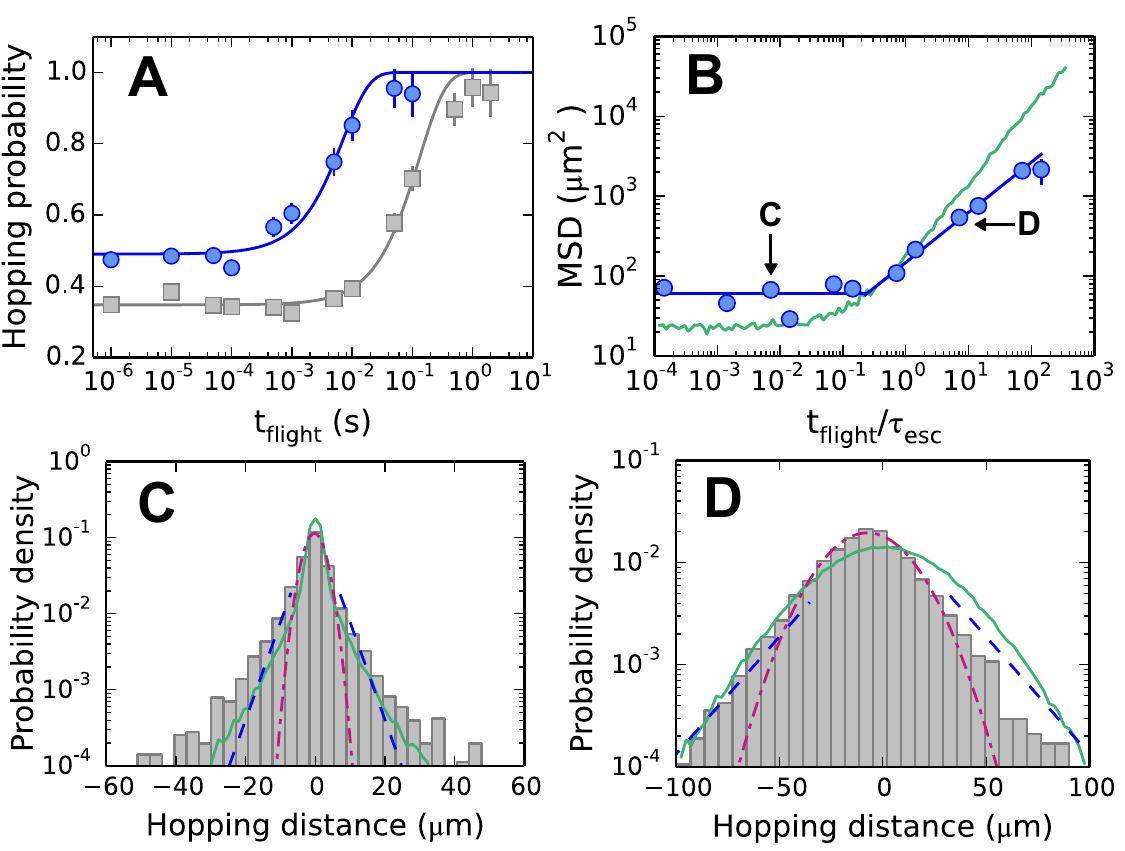}
\caption{(A) Hopping probability versus $t_\mathrm{flight}$ for $U_\mathrm{low}=\SI{210}{\micro \K}$ (blue circles) and \SI{560}{\micro \K} (grey squares).
The solid lines are fits of a cumulated exponential distribution revealing the escape times of $\tau_\mathrm{esc} = \SI{7}{\milli \s}$ and \SI{131}{\milli \s}.
(B) MSD versus $t_\mathrm{flight}$ for a single ramping process.
When $t_\mathrm{flight}$ approaches $\tau_\mathrm{esc}$ the diffusion process starts.
The solid blue line is a double linear fit to the data points and the green solid line is the result of a numerical simulation of the Langevin equation (see SOM). The underlying single step distance distributions are shown for \SI{0.1}{\milli \s} (C) and \SI{50}{\milli \s} (D). 
With increasing flight time the central limit theorem applies and the single step distribution evolves into almost a Gaussian shape. In both images the best fitting Gaussian distribution is depicted (dashed-dotted). The blue dashed line is a fit of $\phi(L)$ (see eq. \ref{eq:L0}) and the green solid line is again the result of the numerical simulation.}
\label{fig:posDist}
\end{figure}
From the data sets, we can either extract ensemble averaged information, by centring all data sets on the first atomic position; this yields MSD values and thus standard diffusion parameters (see Figures \ref{fig:posDist}, \ref{fig:diffusion}).
Alternatively, we can investigate single particle trajectories (see Fig.~\ref{fig:setup} C) from which we can obtain information such as hopping probability (Figure \ref{fig:posDist}A) or hopping distance (Figure \ref{fig:posDist}C, D). We emphasize that, for investigations of ergodicity or correlation functions, it is crucial to have access to the \textit{single} particle trajectories. 
The intra- and inter-well dynamics of the atom in the periodic potential is precisely described by a Langevin equation (see Supplementary Material). In order to analytically describe the essential dynamics in various regimes, we use a continuous-time walk approach (CTRW)\cite{Berkowitz2006, Nelson2001, Scher1973}.
It considers dynamical processes of individual particles as jump processes between lattice sites, neglecting the detailed intra-well dynamics. The waiting time at a certain position as well as the hopping distance in each time step are considered random variables with certain distributions.
The case we consider here realizes such a CTRW with a distribution $\psi(\tau_\mathrm{w})$ of waiting times $\tau_\mathrm{w}$ and distribution $\phi(L)$ of hopping distances $L$ according to
\begin{align}
\label{eq:L0}
\psi(\tau_\mathrm{w}) = \frac{1}{\tau_\mathrm{esc}}\, e^{- \frac{\tau_\mathrm{w}}{\tau_\mathrm{esc}}}, \quad \phi(L) = \frac{1}{2 L_0}\, e^{-\frac{\vert L \vert}{L_0}},
\end{align}
with L0 and $\tau_\mathrm{esc}$ the characteristic length and time scales of the system, which we experimentally control.
We stress that these exponential distributions are not presumed \textit{à priori}, but emerge from the experimental setting. Furthermore, using both the experimentally measured distributions and numerically simulated distributions from a Langevin model, the CTRW \textit{analytically} predicts our findings, such as apparently normal diffusion while the hopping distribution is exponential.
In fact, this specific CTRW describes the rather general case of a diffusing particle coupled to a thermal reservoir while moving in a periodic potential.
Engineering of the ensuing dynamics exploits three mechanisms, which allow us to change the diffusion coefficient by more than three orders of magnitude, as well as changing the MSD from normal diffusive to subdiffusive. First, the optical molasses triggers the diffusion process, as photon scattering leads to random momentum kicks summarized by the diffusion coefficient $D$; furthermore, the associated Doppler cooling force leads to damping with coefficient $\gamma$. Second, the periodic potential adds another time scale $\tau_\mathrm{esc}$ on which an atom can escape its potential well. And third, switching off the molasses intentionally for short periods of time leads to a temperature increase due to phase noise of the optical lattice potential. While this does not alter the steady state temperature in the molasses, it does heat the sample, leaving the atom in a non-equilibrium state (see Supplementary Information for details). 
While the diffusion of atoms in an optical molasses is well known in the context of laser cooling \cite{Bardou1994}, and for subdoppler processes, for instance, Levy flights have been observed \cite{Sagi2012}. For our atomic energies, however, Doppler cooling is the relevant and we use the two other mechanisms to control the diffusion.
The escape time from a potential well is the relevant time scale to describe diffusion dynamics between the lattice wells. For $U \gg k_B T$ the average time for such an escape process is defined by \cite{Kramers1940, Hanggi1990, Ferrando1993}
\label{eq:escTime}
\begin{equation}
\tau_\mathrm{esc} =\frac{\pi k_B T}{4\gamma U} e^{U/k_BT}.
\end{equation}
We measured an escape time of $\tau_\mathrm{esc} = \SI{7}{\milli \s}$ by fitting an accumulated exponential function to the measured hopping probability (Fig. \ref{fig:posDist}A). It is in good agreement with the numerically obtained value of 8 ms from a Langevin simulation. In addition to the waiting time, we have also extracted the distribution of hopping distances, see Figure \ref{fig:posDist}C. It is given by an exponential distribution, yielding a first indication that the central limit theorem for this case does not apply.
Intuitively, the exponential distributions can be understood from the underdamped motion of the atom in a periodic potential. 
An atom will jump whenever it acquires an energy larger than the potential depth from a random fluctuation. 
Since the atoms escape immediately once their energy is large enough, their kinetic energies will be distributed narrowly above zero and be given by the exponential tail of the Boltzmann energy distribution inside the well.
In the underdamped regime, the energy dissipated per traversed well is approximately constant and thus the exponential distribution of energies translates directly into an exponential distribution of flight lengths.
The additional time scale $\tau_\mathrm{esc}$ is directly evident from the measured ensemble averaged dynamics and the MSD observed for a diffusion on a periodic potential, shown in Figure \ref{fig:posDist}B. 
The diffusion effectively starts for $t_\mathrm{flight} > t_\mathrm{esc}$ and is apparently normal with $\delta(t) \propto t^\alpha$ and $\alpha = 1.03$. 
We engineer various diffusion coefficients and different exponents of the time-evolution of the MSD exploiting the third mechanism, namely short heating phases. 
While such heating phases hence affect the atomic dynamics without molasses only marginally, see Supplementary Material, they allow us to transfer the diffusion properties to a broad range of time scales $t_\mathrm{flight}$, see Figure \ref{fig:diffusion}A. The MSD extracted shows diffusion with very different diffusion coefficients ranging over more than three orders of magnitude, see Fig.~\ref{fig:diffusion}B, and MSD exponents ranging from normal diffusion to subdiffusive, see Fig.~\ref{fig:diffusion}C. 
The range of diffusion constants realized can be understood considering the diffusion coefficient as a function of the parameter $t_\mathrm{flight} \cdot \gamma$ (see Figure \ref{fig:diffusion}B), which quantifies the efficiency of molasses cooling during the time of flight. 
Thus the system can be tuned from heating-dominated to cooling-dominated by choosing the proper flight time. 
The range of exponents in the MSD evolution can be tuned by the \textit{absolute} time $t_\mathrm{flight}$ from linear $\alpha =1 $, to subdiffusive $\alpha<1$, see Figure \ref{fig:diffusion}C. 
Intuitively, this can be understood considering the parameter $\sqrt{D\cdot t_\mathrm{flight}}$ determining the average distance which the particle has moved in the weak harmonic potential arising from the dipole trap, and thus defining the characteristic length scale $L_0$ of our CTRW. 
For short flight times, the particle experiences only the slowly varying bottom of the harmonic confinement, while for long flight times the harmonic confinement leads to an increasing repulsive force, slowing down the diffusion for increasing time, i.e.~rendering it sub-diffusive. 
This large range of diffusion parameters allows to trace the behavior of atom diffusion in the subdiffusive regime. 
Thus for our experiment, all parameters of the CTRW can be controlled, $\tau_\mathrm{esc}$ via the external periodic potential, and $L_0=\sqrt{D\cdot t_\mathrm{flight}}$ via $t_\mathrm{flight}$.
\begin{figure}
\includegraphics[width=0.3\textwidth]{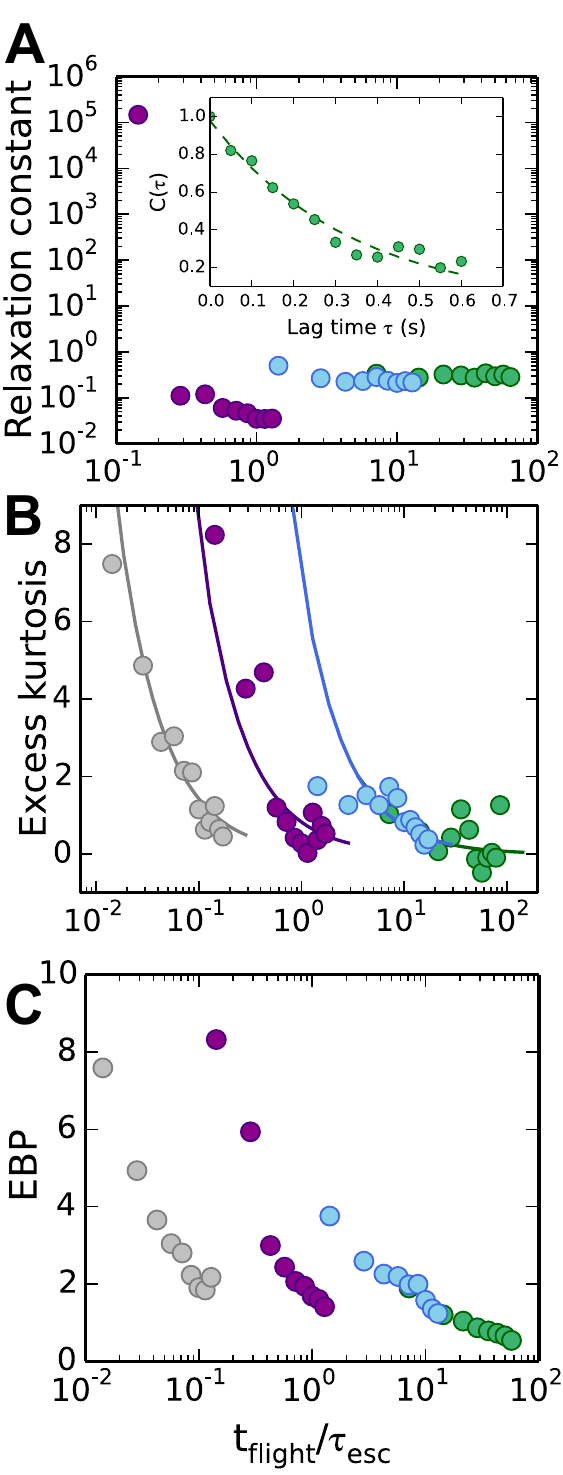}
\caption{(A) Relaxation times obtained from an exponential fit to the position correlation functions. In contrast to the expected correlation function, a decay is observed for longer flight times as shown in the inset for $t_\mathrm{flight}$  = \SI{50}{\milli \s} and the first time step in the sequence.
(B) Excess kurtosis (eq.\ref{eq:kurt}) calculated for the same parameters as before. The solid lines are plots of the analytic expression for the decay of the kurtosis calculated from our CTRW model. Here $\tau_\mathrm{esc}$ is adjusted for each parameter set to fit the calculated values.
(C) Ergodicity breaking parameter calculated for the same parameters as before. Even for flight times 100 times larger than $\tau_\mathrm{esc}$ ergodicity is not yet reached. 
}
\label{fig:ergodicity}
\end{figure}
Importantly, for small $t_\mathrm{flight}$ and thus $\alpha \approx 1$, from our CTRW model the MSD can be analytically derived to evolve according to 
\begin{align}
\langle x^2(t) \rangle = \frac{2 L_0^2}{\tau_0} t, 
\label{diffusion-analytic}
\end{align}
i.e.~showing normal diffusion at all times, which is in excellent quantitative agreement with the experimentally measured diffusion.
In order to probe properties beyond the first moment, we consider the position correlation function of the single-particle trajectories \cite{Fulde1975,Risken1978}
\begin{equation}
C(t,\tau) = \frac{\langle\Delta x(t) \Delta x(t+\tau)\rangle}{\sigma^2(t)},
\end{equation}
where $\sigma$ denotes the standard deviation.
By definition, the value of the correlation function for time lag $\tau =0$ is unity and decays for increasing time lag, i.e.~the positions at increasing time-separations become more and more uncorrelated. 
In steady state it is expected to be unity for all times and time-lags, thus identical to the behavior of the standard Brownian motion. For a subdiffusive system this is only slightly modified, see Fig.~\ref{fig:ergodicity}A.
Thus from these observables, especially for the apparently normal diffusion, even the position correlation function does not yield any sign of non-Gaussian dynamics.

This is markedly changed by considering the step size distribution evaluating individual hopping events. Here, our model predicts an exponential hopping distribution in excellent agreement with the experimental finding (see Figure \ref{fig:posDist}C). This can be quantified by defining the excess kurtosis
\begin{align}
\label{eq:kurt}
\kappa=\frac{\langle \Delta x^4 \rangle}{3 \langle \Delta x^2\rangle^2 -1},
\end{align}
which vanishes for a Gaussian distribution, but is expected to decay in our CTRW $\kappa \propto 2 \tau_\mathrm{esc}/t$. Our data shows both this pronounced non-Brownian distribution (Fig.~\ref{fig:posDist}C) as well as an unexpected large excess kurtosis (see Fig.~\ref{fig:ergodicity}B).

This finding suggests also non-ergodic dynamics on the observed time scales, which does not relax on the time scales $\tau_\mathrm{esc}$ of $\gamma^{-1}$ as expected from Brownian motion.  
From our data, ergodicity can be directly tested by comparing the ensemble average to the time averages of single particle trajectories. 
In order to nonergodicity, we use the ergodicity breaking parameter introduced in \cite{Burov2011}, see Supplementary Material. 
Essentially it is the normalized difference between time- and ensemble averaged MSD in the limit of infinite measurement time. 
For an ergodic system its value is zero, a larger value shows that the system is not ergodic. 
As shown in Figure \ref{fig:ergodicity}C, for all measurement times we find that the system has not reached the ergodic regime. While it is clear that a non-equilibrium system needs a non-zero relaxation time, for Gaussian diffusion it is expected to be established on time scales of the inverse cooling rate $\gamma^{-1}$. In our case, in contrast, we find that ergodicity is not established on time scales large compared to the flight time $t_\mathrm{flight}$ or the damping time $\gamma^{-1}$ when the system is expected to be thermalized. Even for several ten escape times $\tau_\mathrm{esc}$ the system considerably breaks ergodicity. 

Comparing the information obtained from ensemble averages as well as single particle trajectories, our first central result is that, for apparently normal diffusion in a periodic potential, we reproduce and theoretically model the markedly non-Brownian properties of the step size distribution and ergodicity breaking.
Our data is well explained by analytical predictions of a CTRW with exponential distributions emerging from the periodic potential. Moreover, the dynamics is reproduced by numerical simulations of thermally driven diffusion on a periodic potential. The findings resemble the observation of enhanced non-Gaussian diffusion in soft materials \cite{Wang2012}, and our system can serve as a test-bed for future studies of engineered diffusion in, e.g., noisy potentials.
Furthermore, our findings also show that this behavior persists when entering the subdiffusive regime: The step size distribution shows exponentially decaying wings and ergodicity is violated on time scales larger than the characteristic time scales of the problem. 
In contrast, this is neither reflected in the MSD nor the correlation function. 
Our work thus emphasizes the necessity to consider higher order moments of the probability distributions, far beyond the simple MSD and beyond the position correlation function in order to characterize and understand the physics of complex diffusion processes. 

\section*{Supplementary Material}
\subsection*{Trapping and Imaging Single Atoms}
\begin{figure}
\includegraphics[width=0.49\textwidth]{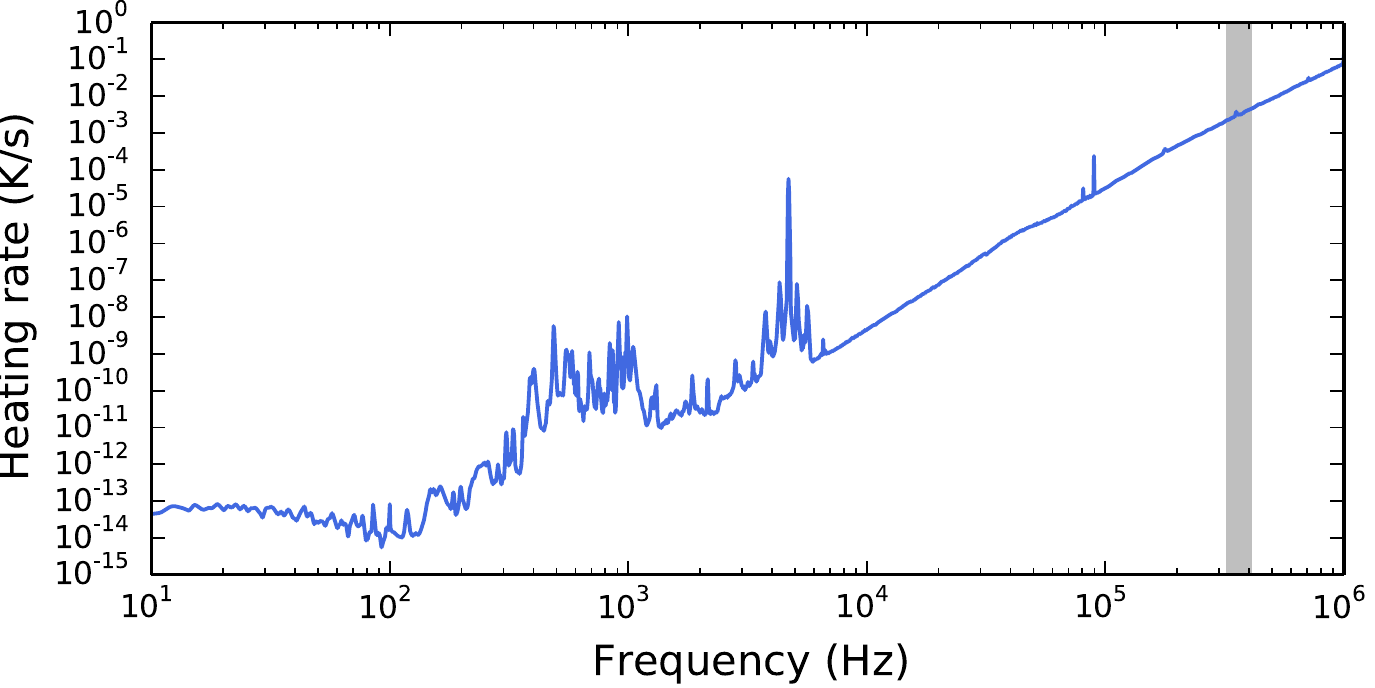}
\caption{Heating rate of the phase noise versus trap frequency of the lattice. The grey shaded area indicates the relevant frequencies for the exponential slope driven in the experiment. 
}
\label{fig:HeatingRate}
\end{figure}

The Caesium atoms are captured in a high gradient ($\approx \SI{250}{G \per \cm}$) magneto-optical trap (MOT).
On average 0-5 atoms are loaded from the background gas and we post-select images, where only one atom is present.
Subsequently we transfer the atoms to a combined optical trap, formed by a running wave optical dipole trap at $\lambda_{DT} = \SI{1064}{\nano \m}$ and an optical lattice formed by two counter propagating laser beams at $\lambda_{Lat} = \SI{790}{\nano\m}$. 
As a consequence the atoms are radially confined by the running wave optical trap with a beam waist $w_{0,DT} = \SI{22}{\micro \m}$ and a power of $\SI{2.6}{W}$ leading to a potential depth of $U_{0,DT} = \SI{1}{\milli \K}$.
Axially the atoms are trapped within the sites of the lattice formed by two counter propagating laser beams at $\lambda_{lat} = \SI{790}{\nano \m}$ with a beam waist of $w_{0,Lat}= \SI{29}{\micro \m}$ and a maximum power of $\SI{650}{\milli \W}$ per beam.
During the experimental sequence only the lattice potential is lowered, while the radial confinement is held constant at all times.
This allows diffusion of the atoms only along the lattice axis and effectively forms a one dimensional system.
\newline
We use fluorescence imaging to precisely count the atom number during the MOT phase as well as to extract the position of the atoms in the lattice.
A high numerical aperture objective (NA=0.36) in distance of $\SI{30.3}{\milli \m}$ to the atoms position collects about 3.3\% of the fluorescence photons.
The collimated light is focussed onto an EMCCD camera (Andor iXon 3 897) with a lens of focal length $f=\SI{1000}{\milli \m}$, yielding a magnification of $M=33$.
Exposure times of $\SI{500}{\milli \s}$ induce a signal to noise ratio of $\approx 5$.
The point spread function of the image system limits the position resolution to an uncertainty of $\SI{2}{\micro \m}$.
For further details of our setup we refer to \cite{Hohmann2015}.
\newline
Jiggling of the position of the potential minima of the lattice is the main source of heating in our system.
The position of the potential minima is dominated by the phase of the two beams of the lattice.
Phase differences are mainly produced by electronic noise of the driver electronics of the acousto optic modulators (AOMs) used for fast switching and phase shifting of the lattice beams.
In axial direction of the lattice the heating rate is given by \cite{Savard1997}:
\begin{equation}
\dot{T} = \frac{4(\pi)^4}{3k_B}m_{Cs}\nu_\mathrm{Lat}^4S_z{\nu_\mathrm{Lat}}
\end{equation} 
where $S_z$ is the noise density and $\nu_{Lat}$ is the lattice frequency.
During the ramp process in the sequence, the atoms are heated as no molasses and hence no cooling is present. 
The total temperature change is calculated from the noise spectrum (see Figure \ref{fig:HeatingRate}) and the residence time at each frequency given by the bandwidth of the timing system running the slope.
Here we use an exponential decaying slope with a duration of \SI{10}{ms} and a bandwidth of \SI{1}{kHz} leading to a temperature increase of about \SI{30}{\micro \K} for typical parameters.
In order to characterize the effect of the heating phases on the diffusion dynamics, we have investigated diffusion in the periodic potential \textit{without} molasses but with a heating period at the beginning of each step. It was positioned in the sequence before the periodic potential was lowered to initialize the dynamics, for a constant flight time of $t_\mathrm{flight}=\SI{10}{\milli\s}$ between two images, see Figure \ref{fig:diffusion}A.
Here, heating of the periodic potential is the only driving process, leading to a much slower diffusion with $D=\SI{460}{\micro \m \squared \per \s}$, because the constant driving due to photon scattering is absent. Hence atoms keep their initial energy throughout the flight time. These heating phase is also evident in the hopping probability, revealed as a non-zero probability offset for small flight times $t_\mathrm{flight}$, see Fig.~\ref{fig:posDist}A.

\subsection*{Quantifying Ergodicity}
The ergodicity breaking parameter (EBP) is defined as \cite{Burov2011}
\begin{equation}
\mathrm{EBP} = \lim\limits_{T \to \infty}\frac{\langle (\bar{\delta^2})^2\rangle - \langle \bar{\delta^2}\rangle^2}{\langle \bar{\delta^2} \rangle ^2}.
\end{equation}
Here $\bar{\delta^2}$ is the mean squared displacement (MSD) and $< >$ denotes the ensemble average.
Therefore the EBP is calculated from the ensemble averaged squared MSD minus the ensemble averaged MSD squared over the ensemble averaged MSD squared. For large enough observation times and normal diffusion this ratio results to zero, as the terms in the nominator approach to the same value.

\subsection*{Control of Diffusion Parameters}
Here the diffusion process is dominated by three main parameters: the diffusion constant D, the damping constant $\gamma$ and the noise term $\xi$.
These parameters can be directly related to our experimental parameters. First of all $\gamma$ is given by the friction coefficient $\beta/m_\mathrm{Cs}$ of the molasses \cite{metcalf2001}
\begin{equation}
\gamma = \beta/m_\mathrm{Cs} = \hbar k^2 \frac{4 s_0 \delta/\Gamma}{m_\mathrm{Cs}(1+ s_0 +(2 \delta/\Gamma)^2)^2}
\end{equation} 
which is for our parameter set in the order of \SI{1e3}{\per \second}. Here $\delta$ is the detuning of the molasses beams frequency, $\Gamma$ the natural line width of the Cesium D2 line transition and $s_0 = I/I_s$ the saturation parameter given by the ratio of molasses intensity $I$ over saturation intensity $I_s$ .
All of these parameters can be easily and precisely controlled in our system.
Directly related to the friction is the diffusion coefficient $D = k_B T\gamma^{-1}$, where T is the atoms temperature mainly set by the molasses or by heating of the lattice due to phase noise.
The randomness in the system is described by Gaussian white noise $\xi$ originating from individual, random photon scattering events at rate $\Gamma_\mathrm{scat}$.

\subsection*{Numerical model}
The observed dynamics in the absence of intentional heating periods is theoretically well described by a Langevin equation assuming a single underdamped Brownian particle in a periodic potential
\begin{equation}
\label{eq:Langevin}
m\dot{v} = -m\gamma v - U k \sin(2kx) + \sqrt{2D}\xi.
\end{equation}
Here, the first and last terms are contributions due to illumination with the optical molasses, i.e.~three pairs of counter-propagating red-detuned, near-resonant laser beams. 
On the one hand, the red-detuning leads to cooling of the atoms and can be described as a classical damping term for a particle with mass $m$ and damping coefficient $\gamma$, arising from the Doppler cooling force.
On the other hand, random absorption and re-emission processes drive the microscopic motion, described by diffusion coefficient $D$ and Gaussian white noise $\xi$, originating from individual, random photon scattering events at rate $\Gamma_\mathrm{scat}$. 
As a consequence the light field acts as a reservoir of constant and adjustable temperature where the atom is coupled to. During the interaction with the light field the atom relaxes to a temperature determined by the molasses on time scales of the inverse cooling rate $\gamma^{-1} = \SI{90}{ms}$.
The center term in equation (\ref{eq:Langevin}) describes the periodic trap discussed above, with depth $U$ and periodicity $d = \lambda/2 = k/(4 \pi)$.
Numerically solving the Langevin equation allows to extract the MSD and the simulated dynamics is in good agreement with our experimental observations.

\subsection*{Continuous time random walk}
While the Langevin dynamics yields good agreement with the experimentally observed motion of the atoms, it is not amenable to analytical treatment and can only be solved through numerical simulation.
For this reason, we now present an even simpler model of the atomic motion in terms of a continuous-time random walk (CTRW).
We will show that this model captures the essential dynamics of the atoms in terms of a jump process between different lattice sites.
Keeping these observations in mind, we define our CTRW model as follows:
We take both the waiting time and jump length distributions to be exponential,
\begin{align}
\psi(\tau_\text{w}) = \frac{1}{\tau_0} e^{-\frac{\tau_\text{w}}{\tau_0}} \qquad \phi(L) = \frac{1}{2 L_0} e^{-\frac{|L|}{L_0}}
\end{align}
The parameters $\tau_0$, the average waiting time, and $L_0$, the average jump length, are obtained from fitting the experimental data with the above distributions, where $\tau_0 = \tau_\mathrm{esc}$, the measured escape time from a potential well.
If up to time $t$, $m$ jumps have occurred, then the total displacement $x(t)$ is the sum over the $m$ corresponding jump lengths $L_{i=1,...,m}$.
The distribution of the total number of lattice sites traversed can be obtained in Fourier-Laplace-space from the Montroll-Weiss equation,
\begin{align}
\hat{\tilde{P}}(k,s) = \frac{1 - \tilde{\psi}(s)}{s} \frac{1}{1-\tilde{\psi}(s) \hat{\phi}(k)},
\end{align}
where $\tilde{\psi}(s)$ is the Laplace transform of the waiting time distribution and $\hat{\phi}(k)$ is the (discrete) Fourier transform of the jump length distribution.
These are given by,
\begin{align}
\tilde{\psi}(s) = \frac{1}{1+\tau_0 s} \qquad \hat{\phi}(k) = \frac{1}{1+k^2 L_0^2} .
\end{align}
The Laplace inversion is readily performed and yields,
\begin{align}
\hat{P}(k,t) = e^{-\frac{t}{\tau_0} (1 - \phi(k))} = e^{-\frac{t}{\tau_0} \frac{k^2 L_0^2}{1+k^2 L_0^2}}. \label{characteristic-function}
\end{align}
For the Fourier inversion there is no closed-form representation, however, we may deduce several properties from the above expression.
Firstly, from the characteristic function Eq.~\eqref{characteristic-function} we can easily obtain the variance via differentiation,
\begin{align}
\langle x^2(t) \rangle = \partial_k^2 \hat{P}(k,t) \Big\vert_{k = 0} = \frac{2 L_0^2}{\tau_0} t. \label{diffusion-analytic}
\end{align}
This corresponds to normal diffusion for all times, with a diffusion coefficient $D = L_0^2/\tau_0$.
Secondly, for short times $t \ll \tau_0$, we can expand Eq.~\eqref{characteristic-function} to find,
\begin{align}
\nonumber
\hat{P}(k,t) \simeq 1 - \frac{k^2 L^2}{1+k^2 L^2} \frac{t}{\tau_0} \quad \Rightarrow \\
\quad P(x,t) \simeq \delta(x)\Big(1-\frac{t}{\tau_0}\Big) + \frac{t}{\tau_0}\frac{1}{2 L_0} e^{-\frac{|x|}{L_0}} .
\end{align}
For short times, the distribution of the displacement is thus given by an initial $\delta$-peak that evolves into an exponential distribution equivalent to the jump length distribution.
Note that the result for the variance \eqref{diffusion-analytic} remains unchanged in the short-time approximation.
Finally, for long times $t \gg \tau_0$, the characteristic function Eq.~\eqref{characteristic-function} is exponentially small except for small $k \lesssim \sqrt{\tau_0/t}/L$, corresponding to $x \gtrsim \sqrt{t/\tau_0} L$.
If $x$ is not too large, we can then perform a saddle-point approximation around $k = 0$ and to find,
\begin{align}
\hat{P}(k,t) \simeq e^{-\frac{t}{\tau_0} k^2 L_0^2} \quad \Rightarrow \quad P(x,t) \simeq \frac{1}{\sqrt{4 \pi L_0^2 \frac{t}{\tau_0}}} e^{-\frac{x^2}{4 L_0^2 \frac{t}{\tau_0}}} .
\end{align}
The displacement distribution is thus a Gaussian in this limit, as would be expected for the diffusive behavior \eqref{diffusion-analytic}.
However, we stress that this approximation breaks down at very large $x \gtrsim \sqrt{t/\tau_0} L$, where the exponential tails prevail.
Nevertheless, the result for the variance \eqref{diffusion-analytic} is valid \textit{exactly for all times}, demonstrating that normal diffusion does not necessarily imply a Gaussian distribution.

\subsection*{Acknowledgements}
This work was partially funded by the EU (ERC Starting Grant Nr. 278208) and Collaborative Project TherMiQ (Grant Agreement 618074) and partially by SFB/TRR49. T. L. acknowledges funding by Carl Zeiss Stiftung, D.M. is a recipient of a DFG-fellowship through the Excellence Initiative by the Graduate School Materials Science in Mainz (GSC 266), F.S. acknowledges funding by Studienstiftung des deutschen Volkes.

\bibliographystyle{Science}
\bibliography{manuscript}{}

\begin{thebibliography}{10}

\bibitem{Frey2005}
E.~Frey, K.~Kroy, {\it Annalen der Physik\/} {\bf 14}, 20 (2005).

\bibitem{Kramers1940}
H.~Kramers, {\it Physica\/} {\bf 7}, 284 (1940).

\bibitem{Jeon2011}
J.~H. Jeon, {\it et~al.\/}, {\it Physical Review Letters\/} {\bf 106}, 048103
  (2011).

\bibitem{Lax2001}
M.~Lax, W.~Cai, M.~Xu, {\it {Random Processes in Physics and Finance}\/}
  (Oxford, 2001).

\bibitem{Kerner2004}
B.~S. Kerner, {\it {The Physics of Traffic - Empirical Freeway Pattern Features
  , Engineering Applications, and Theory}\/} (Springer, Berlin Heidelberg,
  2004), first edn.

\bibitem{Li2010}
T.~Li, S.~Kheifets, D.~Medellin, M.~G. Raizen, {\it Science (New York, N.Y.)\/}
  {\bf 328}, 1673 (2010).

\bibitem{Han2006}
Y.~Han, {\it et~al.\/}, {\it Science\/} {\bf 314}, 626 (2006).

\bibitem{Fakhri2010}
N.~Fakhri, F.~C. MacKintosh, B.~Lounis, L.~Cognet, M.~Pasquali, {\it Science\/}
  {\bf 330}, 1804 (2010).

\bibitem{DAnna2003}
G.~D'Anna, P.~Mayor, a.~Barrat, V.~Loreto, F.~Nori, {\it Nature\/} {\bf 424},
  909 (2003).

\bibitem{Porta2001}
A.~La~Porta, G.~A. Voth, A.~M. Crawford, J.~Alexander, E.~Bodenschatz, {\it
  Nature\/} {\bf 409}, 1017 (2001).

\bibitem{Brokmann2002}
X.~Brokmann, {\it et~al.\/}, {\it Physical Review Letters\/} {\bf 90}, 120601
  (2002).

\bibitem{He2008}
Y.~He, S.~Burov, R.~Metzler, E.~Barkai, {\it Physical Review Letters\/} {\bf
  101}, 058101 (2008).

\bibitem{Manzo2014}
C.~Manzo, {\it et~al.\/}, {\it Physical Review X\/} {\bf 5}, 011021 (2015).

\bibitem{Leptos2009}
K.~C. Leptos, J.~S. Guasto, J.~P. Gollub, A.~I. Pesci, R.~E. Goldstein, {\it
  Physical Review Letters\/} {\bf 103}, 198103 (2009).

\bibitem{Meroz2011}
Y.~Meroz, I.~M. Sokolov, J.~Klafter, {\it Physical Review Letters\/} {\bf 107},
  260601 (2011).

\bibitem{Wang2009}
B.~Wang, S.~M. Anthony, S.~C. Bae, S.~Granick, {\it Proceedings of the National
  Academy of Sciences of the United States of America\/} {\bf 106}, 15160
  (2009).

\bibitem{Wang2012}
B.~Wang, J.~Kuo, S.~C. Bae, S.~Granick, {\it Nature Materials\/} {\bf 11}, 481
  (2012).

\bibitem{Risken1978}
H.~Risken, H.~D. Vollmer, {\it Zeitschrift f{\"{u}}r Physik B Condensed Matter
  and Quanta\/} {\bf 31}, 209 (1978).

\bibitem{Fulde1975}
P.~Fulde, L.~Pietronero, W.~R. Schneider, S.~Str{\"{a}}ssler, {\it Physical
  Review Letters\/} {\bf 35}, 1776 (1975).

\bibitem{viterbi2013}
A.~Viterbi, J.~Omura, {\it Principles of Digital Communication and Coding\/},
  Dover Books on Electrical Engineering (Dover Publications, 2013).

\bibitem{Sancho2004}
J.~M. Sancho, a.~M. Lacasta, K.~Lindenberg, I.~M. Sokolov, a.~H. Romero, {\it
  Physical Review Letters\/} {\bf 92}, 250601 (2004).

\bibitem{Berkowitz2006}
B.~Berkowitz, A.~Cortis, M.~Dentz, H.~Scher, {\it Review of Geophysics\/} {\bf
  RG2003}, 1 (2006).

\bibitem{Nelson2001}
J.~Nelson, S.~Haque, D.~Klug, J.~Durrant, {\it Physical Review B\/} {\bf 63},
  205321 (2001).

\bibitem{Scher1973}
H.~Scher, M.~Lax, {\it Physical Review B\/} {\bf 7}, 4491 (1973).

\bibitem{Bardou1994}
F.~Bardou, J.~Bouchaud, O.~Emile, A.~Aspect, C.~Cohen-Tannoudji, {\it Physical
  Review Letters\/} {\bf 72}, 203 (1994).

\bibitem{Sagi2012}
Y.~Sagi, M.~Brook, I.~Almog, N.~Davidson, {\it Physical Review Letters\/} {\bf
  108}, 093002 (2012).

\bibitem{Hanggi1990}
P.~H{\"{a}}nggi, P.~Talkner, M.~Borkovec, {\it Reviews of Modern Physics\/}
  {\bf 62}, 251 (1990).

\bibitem{Ferrando1993}
R.~Ferrando, R.~Spadacini, G.~E. Tommei, {\it Physical Review E\/} {\bf 48},
  2437 (1993).

\bibitem{Burov2011}
S.~Burov, J.-H. Jeon, R.~Metzler, E.~Barkai, {\it Phys. Chem. Chem. Phys.\/}
  {\bf 13}, 1800 (2011).

\bibitem{Hohmann2015}
M.~Hohmann, {\it et~al.\/}, {\it EPJ Quantum Technology\/} {\bf 2}, 23 (2015).

\bibitem{Savard1997}
T.~Savard, K.~O’Hara, J.~Thomas, {\it Physical Review A\/} {\bf 56}, R1095
  (1997).

\bibitem{metcalf2001}
H.~Metcalf, P.~van~der Straten, {\it {Laser Cooling and Trapping}\/} (Springer
  New York, 2001).

\end{thebibliography}
	
\end{document}